\documentclass{iopart}
\usepackage{graphicx}

\newcommand{\com}[1]{}
\newcommand{\GF}[1]{G_{#1}\left(\mathbf{r};\mathbf{r}'\right)}
\newcommand{\Gbar}[2]{\overline{G}\left(#1,#2,\epsilon\right)}
\newcommand{\G}[3]{G_{#1}\left(#2,#3\right)}
\newcommand{\Pl}[1]{P_l \left( #1 \right)}
\newcommand{\lSum}{\sum_{l=0}^{\infty}}
\newcommand{\Li}[2]{\mathrm{Li}_{#1}\left[ #2 \right]}
\newcommand{\Cut}[1]{}

\newcommand{\EF}[1]{\mathbf{E}_{#1}\left(\mathbf{r};\mathbf{r}'\right)}
\newcommand{\Ebar}[2]{\overline{\mathbf{E}}\left(#1,#2,\epsilon\right)}

\newcommand{\E}[3]{\mathbf{E}_{#1}\left(#2,#3\right)}

\newcommand{\unitvec}[1]{\mathbf{e}_{#1}^{}}

\begin{document}

\title[Dielectric screening in a spherical cavity]{Dielectric
screening in a spherical cavity}
\author{Christopher J Glosser and Roger C Hill}
\address{Department of Physics, Southern Illinois University,
Edwardsville, IL 62026-1654}
\eads{\mailto{cglosse@siue.edu},\mailto{ rhill@siue.edu}}

\begin{abstract}
\\
In this work we examine the electrostatic screening potential due to
a point charge located off-centre in a spherical dielectric cavity.
This potential is expanded for the case in which the dielectric
constant $\epsilon$ is large, several methods of finding the terms in 
the expansion are investigated, and closed-form expressions are found 
through third order in $\epsilon$ along with error bounds.  
Finally, possible uses of these expressions in molecular dynamics 
simulations of isolated charged molecules is discussed. 
\\
\end{abstract}
\submitto{\JPA}

\maketitle
\section{Introduction}


Over the last generation, molecular dynamics simulations have become a vital
theoretical tool in the analysis of the physical interaction of
proteins.  This has been driven in part by the dramatic increase in cheap
computer power in the last decade, which has grown at almost an
exponential rate.

This growth in computer power has resulted in a similarly dramatic
increase in the size of the systems studied utilizing this technique.
What began as a study of 
modest proteins such as myoglobin has expanded to include complicated
systems such as structures embedded in cellular membranes and even a
tobacco mosaic virus \cite{TMV}.

Due to the complexity of these systems, previously ignorable errors
due to approximations in the model are likely to accumulate, 
resulting in inaccurate results and unstable simulations.  Therefore,
it is of vital importance to implement as accurate a representation as
possible, particularly with respect to 
the long-range interactions in the model.
Of these, the most problematic is the electrostatic interaction
between charged elements in the simulation.  In addition to generating
long-range forces, the electrostatic field also polarizes the media in
which the simulation is taking place, effectively creating more
sources for the field in the simulation.  It is this aspect of the
electrostatic interaction that is the most troublesome to implement
accurately while keeping computational time and expense to a minimum.

There have been numerous attempts to circumvent the electrostatics
problem in molecular dynamics models.  The classic way of doing this
is to place the system in a periodic cell and implement Particle Mesh
Ewald dynamics \cite{PME} to account for the long range fields.  This model 
indeed handles the electrostatic problem while keeping the system
size reasonable.  However, it artificially imposes a crystalline
structure on the system which may not be desirable for some applications.  

If one wishes to investigate an isolated structure, then the options
are fairly limited.  A cutoff on the electrostatic interaction is 
usually imposed, but this effectively isolates portions of the
system from one another.  These models also suffer from the defect
that the system in effect becomes finite in size, ignoring a large
portion of the solvent.  Since the solvent --- which is usually water ---
has a large dielectric constant ($\epsilon \approx 80$), it is quite
polarizable.  Therefore, the field generated by the solvent is very
sensitive to the background field.  Multipole methods \cite{Taven}
historically have had some success in dealing with these long range terms.

We wish to reformulate the approach to the electrostatic interaction
in molecular dynamics simulations to take into account this sensitive
dependence of the system on the background field.  Our model should
have the following properties,
\begin{itemize}
\item It should accurately represent the field of the solvent.
\item It should be relatively inexpensive from a computational point of view.
\item The potential in question should be a solution to Poisson's
  equation, so that it represents a physically possible charge distribution.
\end{itemize}

\section{Green's Function for the Screening Potential}

If a charge distribution is placed in a dielectric medium that is
uniform and infinite in extent, the well known result is that the
electric potential is reduced by a factor of $\epsilon$, the relative
permittivity of the dielectric.  This reduction is caused by an
additional ``screening potential'' due to the polarization induced in
the dielectric, which partly cancels the original potential.  The
problem is more complicated when there are dielectric boundaries
involved, as in the case of a charge distribution inside a cavity
within a dielectric.  The interaction of charges embedded in a
dielectric cavity is a surprisingly complicated and rich subject in
the study of classical electromagnetic theory.  Even simple systems
fail to have closed-form solutions for the potential.  If one wishes
to construct a realistic model in which charge interacts with a
dielectric, then some approximation is inevitably necessary.

To begin building our model, let us assume that a dielectric of
uniform relative permittivity $\epsilon$ fills all of space except 
for a spherical cavity centred at the origin, and that the cavity has 
unit radius (i.e.\ all distances are measured in terms of the cavity 
radius).  Suppose that there is a charge distribution $\rho(\mathbf{r})$ 
within the cavity.  The potential everywhere in space may be calculated 
if the Green's function $\GF{}$ is known:
\begin{equation}
  \Phi(\mathbf{r}) =
  \int_{\mathrm{cavity}}\!\! \GF{}
    \rho (\mathbf{r}') \rmd^3 \mathbf{r}'.
\end{equation}
The Green's function satisfies Poisson's equation as a function of 
$\mathbf{r}$ with a unit point charge located at $\mathbf{r}'$ as the 
source:
\begin{equation}
  \nabla^2 \GF{}=-4\pi \delta^3 \left(\mathbf{r}-\mathbf{r}'  \right),
\end{equation}
and must also satisfy the proper boundary conditions on the cavity wall 
and at infinity.  This problem is easily solved by the method of images in 
the limit ($\epsilon \to \infty$), corresponding to a cavity in a conductor.  
However, no closed-form expression for $\GF{}$ is known in terms of elementary 
functions for the case of finite $\epsilon$, except for the trivial case 
in which the charge is located at the centre of the cavity.  It is therefore 
necessary to representing the Green's function with an infinite series or 
other type of approximation.

Let us choose the positive $z$-axis to pass through the source point 
$\mathbf{r}'$, which we are assuming to be within the cavity so that $r' < 1$.  
The classic way of tackling this type of problem is to write 
the Green's function as an infinite series in orthogonal functions, which for 
the case at hand will be Legendre polynomials due to the azimuthal symmetry of 
the problem.  The potential due to the unit point charge alone is
\begin{equation}
  \GF{\mathrm{point}}=\frac{1}{\left| \mathbf{r}-\mathbf{r}'  \right|}
  =\frac{1}{\sqrt{r^2+r'^2-2 r r' \cos \theta}},
\end{equation}
which has a well known expansion in terms of Legendre Polynomials 
\cite{Griffiths,Jackson}.  
For looking at boundary conditions we will be interested in the region near 
the cavity wall, for which $r > r'$ and the expansion is
\begin{equation} \label{point}
  \GF{\mathrm{point}} = \sum_{l=0}^{\infty}\frac{r'^{l}}{r^{l+1}} \,
  \Pl{\cos \theta}.
\end{equation}
There is also a contribution $\GF{\mathrm{screen}}$ to the Green's function, 
the screening potential, due to the polarized dielectric:
\begin{equation}
  \GF{} = \GF{\mathrm{point}}+\GF{\mathrm{screen}}.
\end{equation}
Because the effective polarization charge is only on the surface of the dielectric, 
the screening contribution to the Green's function must satisfy Laplace's equation 
inside and outside the cavity, be finite in each of these regions, and satisfy the 
proper boundary conditions at the cavity wall $r = 1$.  With these conditions in 
mind we can write
\begin{equation} \label{gs1}
  \GF{\mathrm{screen}}
  =
  \left\{ 
    \begin{array}{lr}
      \displaystyle \lSum  A_l r^l \Pl{\cos \theta} & \mbox{for $r<1$},
      \vspace{6pt}\\ 
      \displaystyle \lSum A_l r^{-(l+1)} \Pl{\cos \theta} & \mbox{for $r>1$}.
    \end{array}
  \right.
\end{equation}
The coefficients $A_l$ are the same in both sums to ensure continuity of the 
potential at $r = 1$, which is one of the boundary conditions.

To find the coefficients $A_l$, we enforce the other boundary condition, which 
is that the electric displacement be continuous across the boundary.  This amounts
to a condition on the radial derivative of the complete Green's function:
\begin{equation}
  \left. \frac{\partial G}{\partial r} \right|_{r \to 1-} \!\!
  = \epsilon
    \left. \frac{\partial G}{\partial r} \right|_{r \to 1+}.
\end{equation}
The result is
\begin{equation}
  A_l = -\,\frac{\left(\epsilon-1\right)\left( l+1 \right)}{1+\epsilon
    \left( l+1\right)} \, r'^l.
\end{equation}
Substituting these into \eref{gs1} we find our expression for the
screening Green's function to be
\begin{equation} \label{GGbar}
  \GF{\mathrm{screen}}
  = \left\{ 
    \begin{array}{lr}
      \Gbar{r r'}{\cos \theta}  & \mbox{for $r<1$},
    \vspace{6pt}\\
    \displaystyle\frac{1}{r} \,\, \Gbar{\frac{r'}{r}}{\cos \theta}
      & \mbox{for $r>1$},
    \end{array}
  \right.
\end{equation}
where
\begin{equation} \label{Gseries}
  \Gbar{x}{u} \equiv -\left( \epsilon -1 \right)
    \lSum \frac{l+1}{l+\epsilon \left( l+1 \right)} \, x^l \Pl{u}
\end{equation}
for $0 \le x \le 1$ and $-1 \le u \le 1$.  
This series solution is well known in the literature
\cite{Burko,Messina,Iversen}.  We shall refer to the function $\Gbar{x}{u}$ 
as the ``screening function.''

While Equation \eref{Gseries} provides a perfectly legitimate expression for 
the screening function, it is fairly limited to its usefulness 
in numerical simulations of charges within the cavity.  The rate of 
convergence depends on where the charge is within the cavity, and is 
rather slow unless the charge is close to the origin.  However, we can look for 
methods of summing the series with hopes of obtaining an expression that may be 
more useful in calculations.  The coefficient within the sum in \eref{Gseries} 
can be written as
\begin{equation}
  \frac{l+1}{l+\epsilon(l+1)} = \frac{1}{\epsilon+1}
    +
    \frac{1}{\left( \epsilon+1 \right)^2}
    \cdot \frac{1}{l + \epsilon/(\epsilon+1)},
\end{equation}
separating the sum into two parts:
\begin{equation} \label{Gseries2} \fl
  \Gbar{x}{u}=-\,\frac{\epsilon-1}{\epsilon+1} \left[ \lSum x^l \Pl{u}
    +
    \frac{1}{\epsilon+1} \lSum \frac{1}{l + \epsilon/(\epsilon+1)} \, x^l \Pl{u}
    \right].
\end{equation}
The first is just the generating function for the Legendre polynomials
\cite{Arfken},
\begin{equation}
  \lSum x^l \Pl{u} = \frac{1}{\sqrt{1-2 u x +x^2}}.
\end{equation}
The other sum can be evaluated in several ways, one of the simplest of 
which is to start with the generating function again,
\begin{equation} \label{Pgen2}
  \frac{1}{\sqrt{1-2ut+t^2}} = \lSum t^{l} \Pl{u},
\end{equation}
multiply both sides by $t^{\epsilon/(\epsilon+1)-1}$, and integrate:
\begin{eqnarray} \nonumber
  \int_0^{x} \frac{t^{\epsilon/(\epsilon+1)-1}}{\sqrt{t^2- 2 t u+1}} \, \rmd t
  &=&\int_0^{x}  t^{\epsilon/(\epsilon+1)-1} \lSum t^{l} \Pl{u}\, \rmd t
  \\
  &=&\lSum \frac{x^{l+\epsilon/ (\epsilon+1)}}
    {l + \epsilon/(\epsilon+1)} \, \Pl{u}.
\end{eqnarray} 
This allows us to express the Green's function in terms of an 
integral:%
\footnote{This equation is equivalent to the one appearing in references 
  \cite{Messina, Iversen}.}
\begin{equation} \label{Gint} \fl
  \Gbar{x}{u}=-\,\frac{\epsilon-1}{\epsilon+1} \left[
    \frac{1}{\sqrt{1 - 2ux + x^2}}
    + \frac{x^{-\epsilon/(\epsilon+1)}}{1+\epsilon}
    \int^{x}_0 \frac{t^{-1/(\epsilon+1)}\,\rmd t}{\sqrt{1-2ut+t^2}}
    \right] .
\end{equation}
Expression \eref{Gint} lends itself to the following physical interpretation:  
First of all, in the limit $\epsilon \to \infty$, where the dielectric 
becomes a conductor, only the first term in the square brackets survives.  
For the solution inside the cavity, this corresponds to the familiar 
image point charge outside the cavity, and for the solution outside 
the cavity, this corresponds to putting an image charge at the 
location of the original charge, neutralizing the original charge 
and ``grounding'' the conductor.  Now for finite $\epsilon$, 
the image charge is reduced in magnitude by the factor 
$(\epsilon-1)/(\epsilon+1)$ so that there is a nonzero potential
outside the cavity, but this point charge is not enough to satisfy the 
boundary conditions; an additional {\it line\/} image charge is needed,
stretching from the image point charge to infinity (for the inner solution)
or to the origin (for the outer solution).

\begin{figure}
\includegraphics{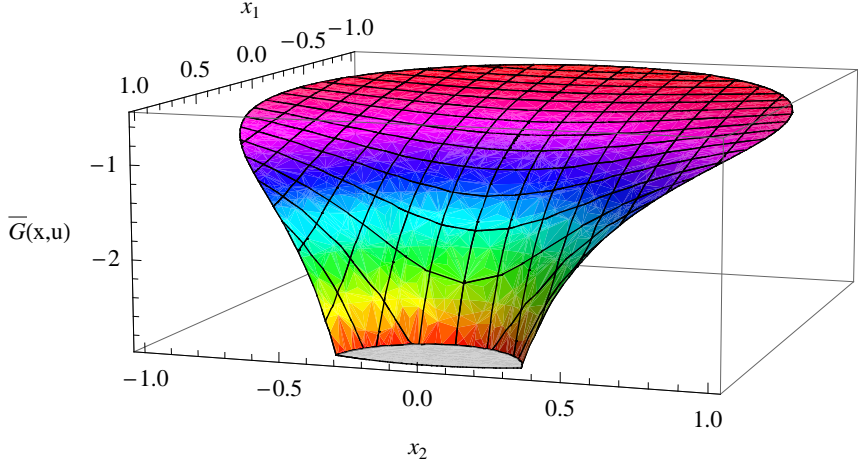}
\caption{A plot of the screening function for $\epsilon = 80$, produced by
  numerically evaluating and plotting \eref{Gint} using the standard packages 
  in {\it Mathematica.\/}  The axis coordinates in this and subsequent surface
  plots were chosen to be $x_1=x u=x \cos \theta$ and 
  $x_2=\pm x \sqrt{1-u^2} = \pm  x \sin \theta$.
}
\label{fullplot}
\end{figure}

\Fref{fullplot} shows a plot of the screening function as a function of 
the variables $x_1 = x\cos\theta$ and $x_2 = x\sin\theta$ in the circular 
region defined by $0 \le x \le 1$.  For points inside the cavity, $x = rr'$ 
as indicated in \eref{GGbar}, so $x_1$ and $x_2$ are proportional to the 
spatial coordinates parallel and perpendicular to the $z$-axis, respectively.  
For a given location $r'$ of the source charge, the cavity boundary 
corresponds to $x = r'$ which is at most one, so a plot of the potential inside 
the cavity may be visualized by truncating the plot in \fref{fullplot} to a 
smaller circular region of radius $r'$, and then expanding the plot to fill a 
region of radius one.  The screening function becomes infinite at the point 
$({x = 1},{u = 1})$, corresponding to the image point charge referred to in 
the previous paragraph, but this singularity only shows up 
in the physical solution when $r' = 1$, i.e.\ when the source charge is 
at the cavity wall.

\section{Expansion for Large Values of the Dielectric Constant}

\Eref{Gint} provides a concise, exact expression for the screening function, 
which can be evaluated numerically and also in terms of Appell hypergeometric 
functions \cite{Appell}.  
However, it is of limited use in actual simulations of molecules in which 
great numbers of these expressions would need to be evaluated.  It is 
therefore useful to look for approximations allowing the use of simpler 
functions.  One such approximation is to recognise that, as we have mentioned, 
the dielectric constant of water is quite large, so that an expansion good for 
large values of $\epsilon$ would be useful.

Based on what we have already derived here, there are two ways in which we can 
obtain a series expansion good for large $\epsilon$.  One way is to go back 
to definition \eref{Gseries} of $\G{n}{x}{u}$ and write it in the form
\begin{eqnarray}
  \G{n}{x}{u} &=& -\,\frac{\epsilon - 1}{\epsilon + 1}
    \lSum \left[1 - \frac{1}{(\epsilon + 1)(l - 1)} \right]^{-1} x^l \Pl{u}
  \nonumber \\
  &=& -\,\frac{\epsilon - 1}{\epsilon + 1}
    \lSum \sum_{n=0}^{\infty} \frac{1}{(\epsilon + 1)^n(l + 1)^n} \, x^l \Pl{u}.
\end{eqnarray}
Reversing the order of summation gives us a series in powers of 
${1/(\epsilon + 1)^n}$:
\begin{equation} \fl
  \Gbar{x}{u} = -\,\frac{\epsilon-1}{\epsilon+1}
    \left(
    \G{0}{x}{u}+
    \frac{\G{1}{x}{u}}{\epsilon+1}+
    \frac{\G{2}{x}{u}}{\left( \epsilon+1 \right)^2}+
    \frac{\G{3}{x}{u}}{\left( \epsilon+1 \right)^3}+
    \dots
    \right), \label{Gseries3}
\end{equation}
where
\begin{equation} \label{Gnseries}
  \G{n}{x}{u} = \lSum \frac{x^l}{(l + 1)^n} \, \Pl{u}.
\end{equation}
The other way to get a series is to expand the integral expression \eref{Gint}, 
noting that
\begin{equation}
  x^{-\epsilon/(\epsilon+1)} t^{-1/(\epsilon+1)}
  = \frac{1}{x} \exp\frac{\ln(x/t)}{\epsilon + 1}
  = \frac{1}{x} \sum_{n=1}^{\infty}
    \frac{\ln^{n-1} (x/t)}{(k-1)! 
    \left( \epsilon+1\right)^{n-1}}.
\end{equation}
This gives the series in \eref{Gseries3} again, where this time the 
coefficients are given by
\begin{equation}
  \G{0}{x}{u}=\frac{1}{\sqrt{1 - 2 u x + x^2}} \label{Gzero}
\end{equation}
and
\begin{equation} \label{Gn}
  \G{n}{x}{u}=\frac{1}{(n-1)!}
    \cdot \frac{1}{x} \int_0^{x}
    \frac{\ln^{n-1} (x/t)}{\sqrt{1 - 2 u t + t^2}} \, \rmd t
    \qquad \mbox{for $n \ge 1$}.
\end{equation}
Expression \eref{Gzero} is clearly equivalent to series expression 
\eref{Gnseries} for $n = 0$, since the former is the generating function 
for the Legendre polynomials.  It is also possible to show directly that 
series expression \eref{Gnseries} and integral expression \eref{Gn} are 
equivalent for $n \ge 0$.  Our purpose in deriving these two expressions 
for $\G{n}{x}{u}$ is that they complement each other:  The series is 
useful for deriving general properties of these coefficients, while the 
integral is useful for doing actual calculations of them.

To get an idea of what sort of functions may be involved in finding 
the coefficients $\G{n}{x}{u}$, let us use the series expression to 
evaluate them in the special case $u = \pm 1$, corresponding to points on 
the $z$-axis.  Since $\Pl{\pm 1} = (\pm 1)^l$, equation \eref{Gnseries} 
gives
\begin{equation} \label{Gnaxis}
  \G{n}{x}{\pm 1} = \lSum (\pm 1)^l \, \frac{x^l}{(l + 1)^n}
  = \pm \frac{1}{x} \Li{n}{\pm x},
\end{equation}
where $\mathrm{Li}_n$ is the polylogarithm function \cite{poly, Dilog, Trilog}:
\begin{equation} \label{Lidef}
  \Li{n}{x} \equiv \sum_{k=1}^{\infty} \frac{x^{k}}{k^{n}}.
\end{equation} 
As can be easily seen from this definition, the polylogarithms satisfy 
a recursion relation,
\begin{equation}\label{Lirec}
  \Li{n}{x} = \int_0^{x} \frac{\Li{n-1}{t}}{t} \, \rmd t.
\end{equation}
The polylogarithm for $n = 0$ is easily found from definition \eref{Lidef}:
\begin{equation}
  \Li{0}{x} = \frac{x}{1-x},
\end{equation}
and those for $n = 1,\,2,\,3,\,\ldots$ may be obtained from this function 
by successive integrations.  As a consequence,
\begin{equation}
  \Li{1}{x} = -\ln (1-x),
\end{equation}
while the dilogarithm $\Li{2}{x}$, the trilogarithm $\Li{3}{x}$, and 
higher-order polylogarithms cannot be expressed in terms of elementary 
functions.  For any positive integer $n$, $\Li{n}{x}$ has a real value 
for $-\infty < x \le 1$, but has a branch cut in the complex plane running 
along the positive real axis from $x = 1$ to $\infty$, across which the
function has a discontinuous imaginary part.

Like the polylogarithms, the coefficients $\G{n}{x}{u}$ also have a similar 
recursion relation which follows immediately from series expression 
\eref{Gnseries}:
\begin{equation} \label{Grec}
  \G{n}{x}{u} = \frac{1}{x} \int_0^x \G{n-1}{t}{u} \, \rmd t,
\end{equation}
and in fact the recursion relation for $x\G{n}{x}{u}$ is the same as for 
$\Li{n}{x}$.  \Eref{Grec} serves as an alternate way to calculate 
$\G{n}{x}{u}$ by starting with $\G{0}{x}{u}$ as given by \eref{Gzero}.

To calculate $\G{n}{x}{u}$ for specific values of $n$, we can either 
use integral expression \eref{Gn} or recursion relation \eref{Grec}; 
both methods seem to lead to about the same degree of complexity.  In 
both methods we have found it useful to make the change of variable
\begin{equation} \label{trans}
  w = \frac{1 + t - \sqrt{1 - 2ut + t^2}}{1 + u}, \quad \mbox{or} \quad
  t=\frac{w}{1 - w} \left( 1 - \frac{1 + u}{2} \, w \right).
\end{equation}
In addition, we define the following variables for the sake of
convenience,
\begin{equation} \label{pq}
  p \equiv \frac{1 + x - \sqrt{1 - 2ux + x^2}}{1 + u}, \qquad
  q \equiv \frac{1 - u}{1 + u}.
\end{equation}
For physical values of $x$ and $u$, the quantities $p$ and $q$ lie in the
range $0 \le p \le 1$ and $0 \le q < \infty$.%
\footnote{The ranges of these variables are interconnected by the fact 
  that, for a given value of $q$, the maximum value of $p$ (corresponding 
  to $x = 1$) is $1 + q - \sqrt{q(1 + q)}$, which is 1 in the limit 
  $u \to \pm1$ but less than 1 otherwise.}
The first-order coefficient $\G{1}{x}{u}$ is then easily evaluated using either 
the integral expression or the recursion relation; both methods lead to 
the same integral:
\begin{eqnarray} \label{G1eval}
  \G{1}{x}{u} &= \nonumber
  \frac{1}{x} \int_0^{x} \frac{\rmd t}{\sqrt{1 - 2ut + t^2}}
  = \frac{1}{x}\int_0^p  \frac{\rmd w}{1 - w}
  \\
  &= -\frac{1}{x} \ln (1 - p)
  = \frac{1}{x} \, \Li{1}{p},
\end{eqnarray}

The higher-order coefficients involve higher-order polylogarithms and 
become increasingly complicated.  A pattern that emerges is that for 
$n \ge 1$, $x\G{n}{x}{u}$ can be expressed entirely in terms of 
logarithms and polylogarithms of rational functions of $p$ and $q$.
Using transformation \eref{trans} on integral \eref{Gn}, 
along with \eref{pq} to eliminate $x$ in the integrand, we obtain
\begin{equation} \label{Ginttrans}
  x\G{n}{x}{u} = \frac{1}{(n - 1)!} \int_0^p
    \ln^{n-1} \left[ \frac{b(1 - w)}{w(1 + q - w)} \right]
    \frac{\rmd w}{1 - w},
\end{equation}
where $b = p(1 + q - p)/(1 - p)$.  Also, from \eref{Grec} we obtain 
a transformed recursion relation,
\begin{equation} \label{Grectrans}
  x\G{n}{x}{u} =
  \frac{1}{x} \int_0^{p} t\G{n-1}{t}{u}
  \left( \frac{1}{w} + \frac{1}{1 - w} - \frac{1}{1 + q - w} \right) \rmd w,
\end{equation}
where $t\G{n-1}{t}{u}$ is assumed to be written in terms of $w$ and $q$.  
Armed with either of these equations we can evaluate the second-order 
coefficient without much difficulty.  For example, using \eref{Grectrans} 
we have
\begin{equation} \fl
  x\G{2}{x}{u} =
  -\int_0^{p} \frac{\ln(1 - w)}{w}\,\rmd w
    - \int_0^{p} \frac{\ln(1 - w)}{1 - w}\,\rmd w
    + \int_0^{p} \frac{\ln(1 - w)}{1 + q - w}\,\rmd w.
\end{equation}
The first integral is just $\Li{2}{p}$ and the second one is easily 
evaluated as $\frac{1}{2} \ln^2 (1 - p)$.  The third integral can be 
found by making the change of variable $w' = q/(1 + q - w)$; the result 
is $-\Li{2}{w'} - \ln(w'/q)$ evaluated at the endpoints.  
The final 
result is
\begin{eqnarray} \label{G2eval} \fl
  \G{2}{x}{u} = \frac{1}{x} \left\{
    \Li{2}{p} + \Li{2}{\frac{q}{1 + q}} - \Li{2}{\frac{q}{1 - p + q}}
    \right.
    \nonumber \\
    \left. {}   
    + \frac{1}{2} \ln^2(1 - p) + \frac{1}{2} \ln^2(1 + q)
    - \frac{1}{2} \ln^2(1 - p + q)
    \right\}.
\end{eqnarray}
The third-order coefficient is considerably more complicated, whether 
we use \eref{Ginttrans} or \eref{Grectrans}.  We evaluated it using 
\eref{Ginttrans}, enlisting the aid of {\it Mathematica\/} 
\cite{Mathematica} to find and manipulate the large number of terms 
and help simplify the expression.  In the process we also made use of 
a number of dilogarithm and trilogarithm identities \cite{Dilog, Trilog}.  
Our result is
\begin{eqnarray} \fl \label{G3eval} \nonumber
  \G{3}{x}{u} = \frac{1}{x} \left\{
    2\,\Li{3}{p}
    - \Li{3}{ \frac{p}{1 + q}}
    + \Li{3}{ 1 - p}
    - \Li{3}{ \frac{1 - p + q}{1 + q}}
    \right.
  \\ \nonumber 
    \left. {}
    - \Li{3}{\frac{1 - p}{1 - p + q}}
    + \Li{3}{ \frac{1}{1 + q}}
    - \Li{3}{ \frac{q}{1 + q}}
    + \Li{3}{ \frac{q}{1 - p + q}}
    \right.
  \\
    \left. {}
    - \Li{3}{ \frac{pq}{1 - p + q}}
    + \ln \frac{(1 - p + q)^2}{(1 - p)(1 + q)^2} \,
      \Li{2}{\frac{q}{1 + q}} 
    \right.
  \\ \nonumber 
    \left. {}
    + \frac{1}{6}\ln^3 \frac{1 - p + q}{(1 - p)(1 + q)}
    - \frac{1}{2} \ln \frac{pq}{1 + q}
      \ln^2 \frac{1 - p + q}{(1 - p)(1 + q)}
    \right.
  \\ \nonumber 
    \left. {}
    - \left[ \ln p \ln (1 - p)
        + \ln(1 + q) \ln \frac{q}{1 + q} \right]
      \ln \frac{1 - p + q}{(1 - p)(1 + q)}
    \right\}.
\end{eqnarray}
One of the complications in writing down the coefficients past 
first order is that there are many different-looking ways to 
express each $\G{n}{x}{u}$, due to the numerous identities 
satisfied by the polylogarithms.  In an attempt to be somewhat 
systematic in our expressions, and also to facilitate the 
study of their behaviour as well as repeated calculations that 
would occur in plotting or numerical simulations, 
we have used these identities where necessary to express the 
results in a form where the arguments of all polylogarithms are 
between 0 and 1 inclusive.

As a check on these results, we can look at their behaviour in 
the limit $u \to 1$, which corresponds to
\begin{equation}
  p \to x,  \qquad  q \to 0.
\end{equation}
In \eref{G2eval}, the first term in the curly brackets becomes 
$\Li{2}{x}$ while the other terms either become zero or cancel 
in pairs, so that $\G{2}{x}{1} = \Li{2}{x}/x$ as expected 
from \eref{Gnaxis}.  In \eref{G3eval}, the first two terms 
give $\Li{3}{x}$ while the rest becomes zero, giving the 
expected result $\G{3}{x}{1} = \Li{3}{x}/x$.  The behaviour 
in the limit $u \to -1$, while correct, is less straightforward; 
it corresponds to
\begin{equation}
  p \to \frac{x}{1 + x},  \qquad  q \to \infty,
\end{equation}
and in this limit we obtain
\begin{equation}
  \G{2}{x}{-1} = \frac{1}{x} \left\{ \Li{2}{\frac{x}{1 + x}}
    + \frac{1}{2} \ln^2 (1 + x)
    \right\},
\end{equation}
which reduces to the expected $-\Li{2}{-x}/x$ by way of one 
of the dilog identities.  The behaviour of the third-order 
coefficient is similar.

It should be pointed out that despite the factor of $1/x$ 
occurring in these expressions, $\G{n}{x}{u}$ is not singular 
at $x = 0$, because the rest of the expression becomes zero 
there.  In fact,
\begin{equation} \label{Gnorigin}
  \G{n}{0}{u} = 1,
\end{equation}
as can be seen from series expression \eref{Gnseries}.

\begin{figure}
\includegraphics[scale=.55]{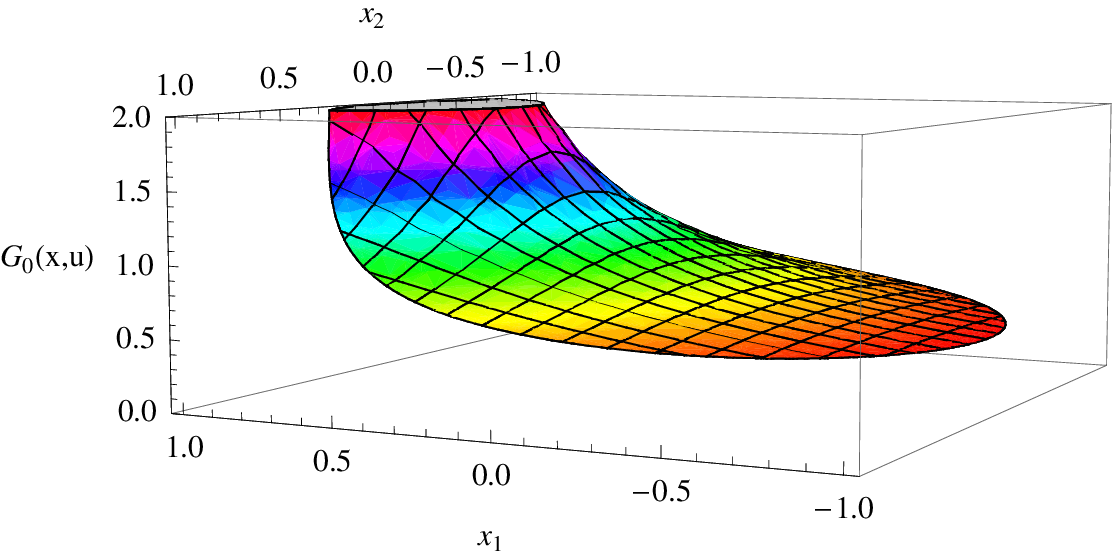}
\includegraphics[scale=.55]{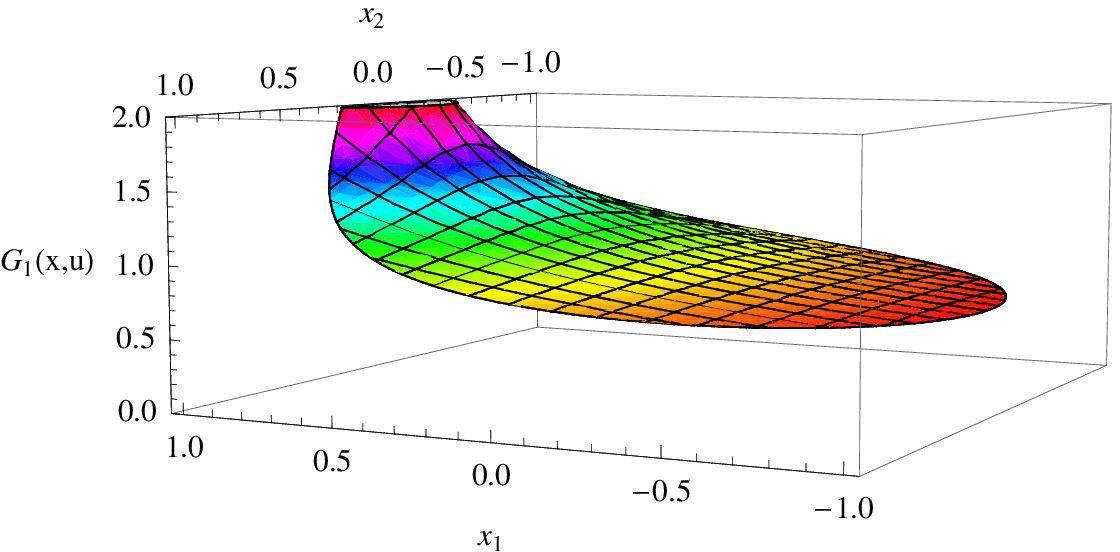}
\\
\includegraphics[scale=.55]{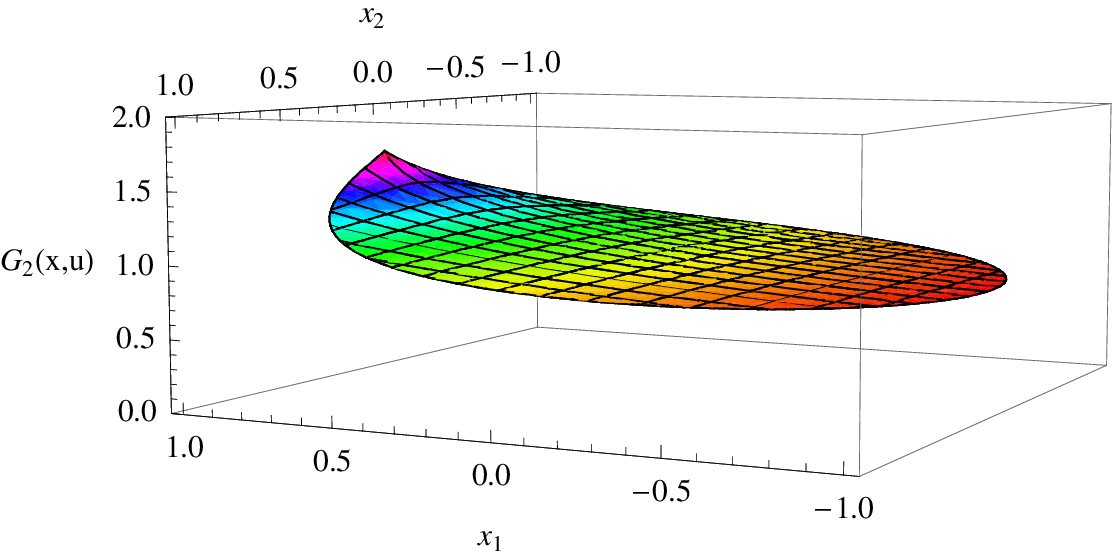}
\includegraphics[scale=.55]{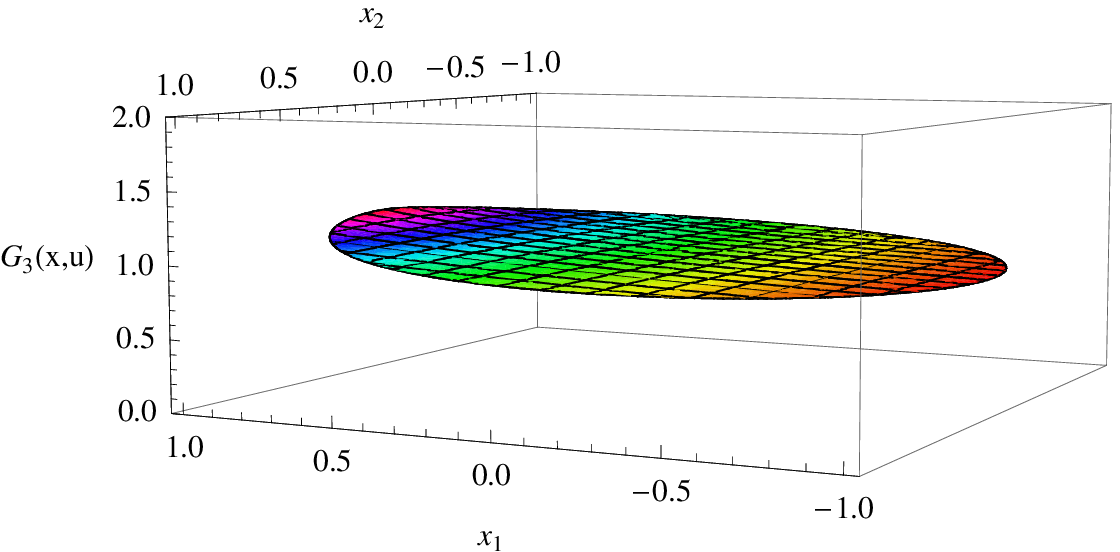}
\caption{The coefficient functions $G_0$ through $G_3$.}
\label{Giplots}
\end{figure}

\Fref{Giplots}
shows three-dimensional plots of $\G{0}{x}{\cos\theta}$ through 
$\G{3}{x}{\cos\theta}$ as functions of $x_1=x\cos\theta$ and 
$x_2=x\sin\theta$, similarly to \fref{fullplot}.  
The
zeroth-order coefficient has an inverse first-power singularity 
at the point $(x = 1, u = 1)$ corresponding to the image 
point charge as per the discussion following equation \eref{Gint}, 
and the first-order coefficient has a logarithmic infinity at 
that point, while the rest of the coefficients are finite 
everywhere in the region.  Despite the increasing complexity 
of the expressions as the order becomes higher, their actual 
behaviour becomes increasingly simple; the plot of the 
third-order term is comparatively flat.
We can in fact obtain bounds on $\G{n}{x}{u}$ in general by noting 
from integral expression \eref{Gn} that for any given $x$, the
integral will have a maximum value when $u = 1$ and a minimum value 
when $u = -1$.  It therefore follows from \eref{Gnaxis} that for 
given $x$, the maximum value is $\Li{n}{x}/x$ and the minimum 
value is $-\Li{n}{-x}/x$.  Since for $0 \le x \le 1$, 
$\Li{n}{x}/x$ and $-\Li{n}{-x}/x$ are strictly increasing and 
decreasing functions of $x$ respectively,%
\footnote{This can be seen by looking at their derivatives, 
using the power series in \eref{Gnaxis}.}
it follows that (for 
$n \ge 1$) the maximum value over all $x$ and $u$ is 
$\Li{n}{1} = \zeta(n)$, 
and the minimum value is 
$-\Li{n}{-1} = (1 - 2^{-n+1})\,\zeta(n)$,
where $\zeta(n)$ is the Riemann zeta function:
\begin{equation}
  \zeta(n) \equiv \frac{1}{1^n} + \frac{1}{2^n} + \frac{1}{3^n}
    + \cdots \,.
\end{equation}
It follows that
\begin{equation}\label{Gnbounds}
  (1 - 2^{-n+1})\,\zeta(n) \le \G{n}{x}{u} \le \zeta(n)
\end{equation}
for all $x$ and $u$ in the physical region, and for $n \ge 1$.  
The $n = 0$ case is easily treated; the lower bound is 
$\G{0}{1}{-1} = \frac{1}{2}$.  The results are numerically, 
\begin{eqnarray} \label{Gnboundsnum}
  0.500000 &\le \G{0}{x}{u} < \infty \nonumber \\
  0.693147 &\le \G{1}{x}{u} < \infty \nonumber \\
  0.822467 &\le \G{2}{x}{u} \le 1.644934  \\
  0.901543 &\le \G{3}{x}{u} \le 1.202057\,. \nonumber
\end{eqnarray}
As $n$ becomes large, the lower and upper bounds both 
approach 1, leading to an increasingly flat plot.  
The coefficient $\G{n}{x}{u}$ could then be approximated by a 
simple polynomial, or even a constant, greatly reducing time in 
computationally intensive problems.  However, for problems 
occurring in practice with large values of the dielectric 
constant, it may not be necessary to keep very many 
terms anyway.  We will look at the errors due to truncating the series
in \sref{sec:err}.

\section{Electric Fields}

While the potential is useful for calculating the potential energy of 
the system, one also needs to be able to calculate the force 
of the various objects upon one another so that the system may evolve
from one time step to the next.  This in turn requires knowing the 
electric field due to the charges and the dielectric.  Corresponding 
to the Green's function $\GF{}$, which is the potential at $\mathbf{r}$ 
due to a unit point charge at $\mathbf{r'}$ in the presence of the 
dielectric, let us define $\EF{}$ be the electric field at $\mathbf{r}$ 
due to a unit point charge at $\mathbf{r'}$ in the presence of the 
dielectric.  The electric field is determined from the Green's 
function by
\begin{equation}
  \EF{} = -\nabla_{\mathbf{r}} \GF{},
\end{equation}
where $\nabla_{\mathbf{r}}$ means the gradient with respect to the 
vector $\mathbf{r}$, keeping $\mathbf{r'}$ constant.  Like the 
potential, the electric field separates into the field to the point 
charge alone plus the ``screening field'' due to the polarized 
dielectric:
\begin{equation}
  \EF{} = \EF{\mathrm{point}} + \EF{\mathrm{screen}}.
\end{equation}
The field $\EF{\mathrm{point}}$ is the familiar Coulomb field,
\begin{equation}
  \EF{\mathrm{point}} =
  \frac{\mathbf{r} -\mathbf{r'}}{|\mathbf{r} -\mathbf{r'}|^3}
  = \frac{(r - r'\cos\theta) \, \unitvec{r}
    + (r'\sin\theta) \, \unitvec{\theta}}
    {(r^2 + r'^2 - 2rr'\cos\theta)^{3/2}},
\end{equation}
where $\unitvec{r}$ and $\unitvec{\theta}$ are unit vectors in 
the direction of increasing $r$ and $\theta$.  
In \eref{GGbar} we expressed the screening part of the Green's function 
inside and outside of the cavity in terms of a single function 
$\Gbar{x}{u}$, which can be thought of as a function of a three-dimensional 
coordinate $\mathbf{x}$ with spherical coordinates $(x,\theta,\phi)$.  
Let us define $\Ebar{x}{u}$ to be the negative of the gradient of this 
function:
\begin{eqnarray} \label{defEbar}
  \Ebar{x}{u} &=& -\nabla_{\mathbf{x}} \Gbar{x}{u}
   = -\left( \unitvec{x} \frac{\partial}{\partial x}
      + \unitvec{\theta} \frac{1}{x}\,\frac{\partial}{\partial\theta} \right)
      \Gbar{x}{\cos\theta} \nonumber \\
   &=& \left(-\,\unitvec{x} \frac{\partial}{\partial x}
      + \unitvec{\theta} \frac{\sqrt{1 - u^2}}{x}\,\frac{\partial}{\partial u}
      \right) \Gbar{x}{u}.
\end{eqnarray}
Then from \eref{GGbar}, the screening field can be written as
\begin{equation} \label{EEbar}
\fl  \EF{\mathrm{screen}}
  = \left\{ 
    \begin{array}{lr}
      r' \, \Ebar{r'r}{\cos\theta} & \mbox{for $r<1$},
    \vspace{6pt}\\
    \displaystyle\frac{r'}{r^3} \, \Ebar{\frac{r'}{r}}{\cos\theta}
      + \frac{1}{r^2} \, \Gbar{\frac{r'}{r}}{\cos\theta} \unitvec{r}
      & \mbox{for $r>1$}.
    \end{array}
  \right.
\end{equation}

The expansion of $\Gbar{x}{u}$ in powers of ${1/(1 + \epsilon)}$ leads to a 
corresponding expansion of $\Ebar{x}{u}$:
\begin{equation} \fl
  \Ebar{x}{u} = -\,\frac{\epsilon-1}{\epsilon+1}
    \left(
    \E{0}{x}{u}+
    \frac{\E{1}{x}{u}}{\epsilon+1}+
    \frac{\E{2}{x}{u}}{\left( \epsilon+1 \right)^2}+
    \frac{\E{3}{x}{u}}{\left( \epsilon+1 \right)^3}+
    \dots
    \right),
\end{equation}
where
\begin{equation}
  \E{n}{x}{u} = -\nabla_{\mathbf{x}} \G{n}{x}{u},
\end{equation}
The zero-order term is found in a straightforward manner, and is just the 
Coulomb electric field due to a unit point charge located on the $x$-axis 
at unit distance from the origin:
\begin{equation}
  \E{0}{x}{u} = \frac{(x - u)\,\unitvec{x} + \sqrt{1 - u^2}\,\unitvec{\theta}}
    {(1 - 2ux + x^2)^{3/2}}.
\end{equation}
For the higher orders, we can avoid having to take some of the derivatives 
of $\G{n}{x}{u}$ explicitly by noticing that if we differentiate both sides of 
recursion relation \eref{Grec} with respect to $x$, we obtain
\begin{equation}
  \frac{\partial G_n}{\partial x}
  = \frac{1}{x} (G_{n-1} - G_n)
    \quad \mbox{for $n \ge 1$}.
\end{equation}
The derivative with respect to $u$ is not so easily found, but even it can be 
simplified.  For $n \ge 1$, we found that our expressions for $\G{n}{x}{u}$ 
were of the form $1/x$ times a function consisting of logarithms and 
polylogarithms of $p$ and $q$.  Therefore, for $n \ge 1$ it is convenient to write
\begin{equation} \label{dGndu}
  \frac{\partial G_n}{\partial u}
  = \frac{1}{x} \, \frac{\partial (xG_n)}{\partial u}
  = \frac{1}{x} \left[ \frac{\partial p}{\partial u} \, \frac{\partial (xG_n)}{\partial p}
    + \frac{\partial q}{\partial u} \, \frac{\partial (xG_n)}{\partial q} \right].
\end{equation}
On the other hand we also know that, again from the recursion relation,
\begin{equation} \label{dxGndx}
  G_{n-1} = \frac{\partial (xG_n)}{\partial x}
  = \frac{\partial p}{\partial x} \, \frac{\partial (xG_n)}{\partial p}
    + \frac{\partial q}{\partial x} \, \frac{\partial (xG_n)}{\partial q}.
\end{equation}
From definitions \eref{pq} of $p$ and $q$, the partial derivatives of them 
with respect to $x$ and $p$ are
\begin{equation} \label{partials}
  \begin{array}{lr}
    \displaystyle \frac{\partial p}{\partial x}
      = \frac{(1 - p)^2 (1 + q)}{(1 - p)^2 + q} ,
    &\quad \displaystyle \frac{\partial p}{\partial u}
      = \frac{p^2 (1 - p) (1 + q)}{2 \left[(1 - p)^2 + q\right]} ,
    \vspace{3pt}\\ 
    \displaystyle \frac{\partial q}{\partial x} 
      = 0 ,
    &\quad \displaystyle \frac{\partial q}{\partial u}
      = -\,\frac{(1 + q)^2 }{2} .
    \end{array}
\end{equation}
Since $\partial q/\partial x = 0$, equation \eref{dxGndx} allows us to 
express $\partial (xG_n)/\partial x$ in terms of $G_{n-1}$, which can then 
be used in \eref{dGndu}, giving
\begin{eqnarray} \nonumber
  \frac{\partial G_n}{\partial u}
 & = \frac{1}{x} \left[
    \frac{\partial p/\partial u}{\partial p/\partial x} \, G_{n-1}
    + \frac{\partial q}{\partial u} \, \frac{\partial (xG_n)}{\partial q} \right] \\  \label{dGndusimp} 
  &= \frac{1}{x} \left[
    \frac{p^2}{2(1 - p)} \, G_{n-1}
    - \frac{(1 + q)^2 }{2} \, \frac{\partial (xG_n)}{\partial q} \right]
\end{eqnarray}
Using these results in \eref{defEbar} along with the fact that 
$\sqrt{1 - u^2} = 2\sqrt{q}/(1 + q)$, we obtain for the $n^{\mathrm{th}}$-order 
coefficient in the electric field
\begin{equation} \label{En} \fl
  \E{n}{x}{u} = \frac{1}{x} (G_n - G_{n-1})\,\unitvec{x}
    + \frac{\sqrt{q}}{x^2} \left[ \frac{p^2}{(1 - p)(1 + q)}\,G_{n-1}
      - (1 + q)\,\frac{\partial (xG_n)}{\partial q} \right]\,\unitvec{\theta}.
\end{equation}
so that we only need to take derivatives with respect to $q$.  
It should be noted that, like $\G{n}{x}{u}$, $\E{n}{x}{u}$ is not singular at 
$x = 0$ despite the inverse powers of $x$, as they multiply factors which 
become zero there.  In simplifying expressions for the potential and electric 
field it is sometimes convenient to use the expressions for $x$ and $G_0$ in 
terms of $p$ and $q$:
\begin{equation}
  x = \frac{p(1 - p + q)}{(1 - p)(1 + q)},
\end{equation}
\begin{equation}
  \G{0}{x}{u} = \frac{1}{\sqrt{1 - 2ux + x^2}}
  = \frac{(1 - p)(1 + q)}{(1 - p)^2 + q}.
\end{equation}

As an example in finding a higher-order coefficient in the electric field, 
let us evaluate $\E{1}{x}{u}$.  From \eref{G1eval}, $xG_1 = -\ln(1 - p)$ which 
is independent of $q$, so no derivatives have to be taken at all in 
evaluating \eref{En}.  After some simplification we obtain
\begin{equation} \fl
  \E{1}{x}{u}
  = -\,\frac{1}{x^2} \left[ \frac{p(1 - p + q)}{(1 - p)^2 + q} + \ln(1 - p)
      \right] \, \unitvec{x}
    + \frac{1}{x^2}\,\frac{p^2 \sqrt{q}}{(1 - p)^2 + q} \, \unitvec{\theta}.
\end{equation}
The $(1 - p)^2 + q$ in the denominator will be recognized as the inverse 
first-power singularity characteristic of $\G{0}{x}{u}$. As with the
potential, the electric field is finite at $x=0$ despite the inverse
powers of $x$.

\section{Analysis of Errors} \label{sec:err}

In implementing our series expansion \eref{Gseries3} for any
particular problem, we need to know when we can when we can truncate
the series for a given desired degree of accuracy (particularly
considering how increasingly complicated the series coefficients become
with each subsequent order).  Suppose that we have evaluated the
series from $n=0$ to $n=N$.  The error $e_N^{}$ due to ignoring the 
terms from $N+1$ to infinity is
\begin{equation}
  e_N^{} = \frac{\epsilon - 1}{\epsilon + 1}
    \sum_{n=N+1}^{\infty} \frac{\G{n}{x}{u}}{(\epsilon + 1)^n}.
\end{equation}
The terms in this series are all positive, and the maximum value of 
$\G{n}{x}{u}$ is $\zeta(n)$ from \eref{Gnbounds}, so
\begin{equation} \label{errnbounds}
  e_N^{}
  \le \frac{\epsilon - 1}{\epsilon + 1}
    \sum_{n=N+1}^{\infty} \frac{\zeta(n)}{(\epsilon + 1)^n}
\end{equation}
This sum can be evaluated numerically for any given $\epsilon$ 
and $N$.  For example, using $\epsilon = 80$ we find
\begin{eqnarray} \label{errornum}
  e_1^{} &\le 2.468\times10^{-4} \nonumber \\
  e_2^{} &\le 2.231\times10^{-6} \\
  e_3^{} &\le 2.482\times10^{-8}. \nonumber \\
\end{eqnarray}
We can also find an upper bound for the {\it fractional\/} error 
$e_N^{}/|\Gbar{x}{u}|$ by noting that the true value 
of $|\Gbar{x}{u}|$ is greater than the $N^{\mathrm{th}}$-order 
calculated value, and therefore the 
fractional error will be less than that found by dividing the error 
by the calculated value.  Also, a lower bound over all $x$ and $u$ 
for the calculated value can be found using the lower bounds for 
$\G{n}{x}{u}$ in \eref{Gnboundsnum}.  For example, suppose that we 
have kept only the $n = 0$ and $n = 1$ terms, again assuming 
$\epsilon = 80$.  The maximum error is $e_1^{}$ in \eref{errornum}, 
while the minimum value of $|\Gbar{x}{u}|$ is at least
\begin{equation}
  |\Gbar{x}{u}|_{\mathrm{min}} = \frac{79}{81}\left(
    0.5000 + \frac{0.6931}{80} \right)
  = 0.4961
\end{equation}
and so an upper bound for the fractional error for any $x$ and $u$ is 
$2.468\times10^{-4}/0.4961 = 4.974\times10^{-4}$.  Similarly, for a 
calculation through second order the fractional error will be less 
than $4.496\times10^{-6}$, and for a calculation through third 
order, less than $5.001\times10^{-8}$.  
Of course, the fractional errors will be much smaller than these 
bounds in regions where $\overline G$ is large, even though the 
error itself do not vary much over the whole region.

\section{Use in Molecular Dynamics Simulations}

Having obtained our expression for the potential to third order in 
$\left( 1+\epsilon \right)^{-1}$, we now turn to the question of
feasibility.  That is, to what degree is this calculation useful to
those doing large-scale simulations of biological systems?

A cursory objection may be that the computation of special functions
as required by this potential 
for each of the charged objects in a simulation would be intractable,
and that it would lead to an inexorable increase in computation
time.  While certainly true, the problem is not as severe as it first
seems.

First, the computation of polylogarithms is very well understood.  The
Taylor series converges rapidly for arguments in the range $0 < x
<1$.  Moreover, a partial fraction technique exists \cite{Pfrac} that
converges extremely rapidly.   For instance, an evaluation of
$\Li{2}{\frac{1}{2}}$ converges to within  $1\times 10^{-7}$ of the
actual value utilizing only 10 divisions.

Second, if the system is of a suitable size, some fields can be
calculated using the total charges of extended structures, such as the
residues comprising the proteins, rather than the constituent atoms.
Since each term in the potential is separately a solution to Laplace's
equation, they each comprise a complete solution.  For instance, one could
approach a large system by calculating the atomic charges exactly
to $O\left[ (1+\epsilon)^{-2} \right]$, while using the $\G{2}{x}{u}$
term to calculate screening by using total charge of the protein
residues and treating them as point charges.

\section{Summary}

For the potential of a point charge in a spherical dielectric cavity,
the authors have computed integral expressions that lead to a
calculable  expansion in inverse powers of the relative permittivity.  The main
feature of this expansion is that the truncated series yields a
potential that is accurate over the physical region while
remaining a solution to Poisson's equation at each order.
For the case of water, truncating the
series to second order leads to a fractional error of no more than 
$4.5 \times 10^{-6}$, and truncating the series to third order leads 
to a fractional error of no more than $5.0 \times 10^{-8}$.

While much of the theoretical groundwork in this article is complete,
the feasibility of this model in an actual simulation environment has
not been studied.  The polylogarithms introduced to describe the
series coefficients $\G{n}{x}{u}$ may be computed rapidly and
precisely using the partial fraction technique, but 
it remains to be seen whether the accuracy gains surmount the time lag
introduced by having additional terms in the potential.  A systematic
study of this would need to be done by altering one of the existing
molecular dynamics packages such as {\it CHARMM\/} \cite{CHARMM} or 
{\it NAMD\/} \cite{NAMD}.

\appendix

\section{Contour Representations}

In addition to the methods that we have presented for finding the 
screening function and the coefficients in the expansion for large 
$\epsilon$, there are other methods using complex-variable techniques.  
For example, the second sum in \eref{Gseries2} can be found by looking 
at the special case $u = 1$, which can be expressed as a hypergeometric 
function \cite{Arfken}:
\begin{equation}
  \frac{1}{\epsilon+1} \lSum \frac{1}{\displaystyle
    l+\frac{\epsilon}{\epsilon+1}}x^l
  =\frac{1}{\epsilon}\, {}_2F_1\left[
    \frac{\epsilon}{\epsilon+1},1;1+\frac{\epsilon}{\epsilon+1};x \right].
\end{equation}
The Legendre polynomials can then be brought in by using a 
contour-integral expression for them,
\begin{equation} \label{Plcontourint}
  \Pl{u} = \frac{1}{2\pi \rmi} \oint
    \frac{z^{-l}}{\sqrt{1-2uz+z^2}} \, \frac{\rmd z}{z},
\end{equation}
where the integration is counterclockwise along any path that encircles 
the origin but does not encircle the branch points of the square root  
(e.g.\ a circle of radius less than 1).
The sum with the Legendre polynomials can then be expressed in terms of 
a contour integral:
\begin{eqnarray} \fl
  \lSum \frac{x^l}{\displaystyle
    l+\frac{\epsilon}{\epsilon+1}} \Pl{u}
  &=&
  \frac{1}{2\pi \rmi}
    \oint \lSum\frac{1}{\displaystyle 
    l+\frac{\epsilon}{\epsilon+1}}
    \frac{\left( z/x \right)^{-l}}{\sqrt{1-2uz+z^2}}
   \, \frac{\rmd z}{z}, \nonumber
  \\
  &=&
  \frac{1+\epsilon}{\epsilon} 
    \frac{1}{2\pi \rmi}
    \oint 
    \frac{
    {}_2F_1
    \left[\frac{\epsilon}{\epsilon+1},1;1+\frac{\epsilon}{\epsilon+1};\frac{x}{z}
    \right]}
    {\sqrt{1-2uz+z^2}}
    \, \frac{\rmd z}{z}. 
\end{eqnarray}
The screening function then becomes
\begin{equation} \label{Gcontour} \fl
  \Gbar{x}{u} =
  -\frac{\epsilon-1}{\epsilon+1} \left[
    \frac{1}{\sqrt{1-2 u x+x^2}}
    +\frac{1}{\epsilon}
    \frac{1}{2\pi \rmi}\oint
    \frac{ 
    {}_2F_1
    \left[\frac{\epsilon}{\epsilon+1},1;1+\frac{\epsilon}{\epsilon+1};\frac{x}{z}
    \right]}
    {\sqrt{1-2uz+z^2}}
    \, \frac{\rmd z}{z},
    \right].
\end{equation}

A similar technique can be used in expressing the $n^{\mathrm{th}}$-order 
coefficient $\G{n}{x}{u}$, using series expression \eref{Gnaxis} for the 
case $u = 1$ in combination with contour integral \eref{Plcontourint}, giving
\begin{equation} \label{Gncon}
  \G{n} {x}{u}
  = \frac{1}{x} \frac{1}{2\pi \rmi}
    \oint\frac{\Li{n}{\displaystyle \frac{x}{z}}}{\sqrt{1-2z u+z^2}} \, \rmd z.
\end{equation}
All of the expressions for the $G_n$ listed in the main body of the
text may be obtained  by
directly evaluating this contour expression.  In particular, it is easy to
derive the recursion relation \eref{Grec} by direct differentiation of
\eref{Gncon}.

\section*{References}
\bibliography{screening_paper_iop}
\end{document}